# Rights by Architecture: A Human-Compatible Sociotechnical Layer for Digital Protection Across Regulatory Regimes

*Completed Research Paper*

**Soheil Human**
Vienna University of Economics and Business, Vienna, Austria
IT:U Interdisciplinary Transformation University Austria, Linz, Austria
Delft University of Technology, Delft, The Netherlands
soheil.human@wu.ac.at

## Abstract

*Digital rights increasingly exist in law but remain difficult to exercise through the information systems that mediate them. Using disciplined conceptual synthesis and problematization, this critical-conceptual IS paper explains the gap through the interaction of legal heterogeneity, conflicting organizational and commercial incentives, fragmented architectures, and asymmetrical control over rights-relevant acts. It then theorizes a human-compatible rights layer: a governed sociotechnical capability for standardized, machine-readable, bidirectional, and jurisdictionally plural communication of requests, consent, refusal, withdrawal, objection, records, and support. Comparing California-style opt-out signals, the EU's mixed lawful-basis regime, and P3P, DNT, GPC, and ADPC, the paper derives seven normative requirements, develops rights by architecture as a bounded emancipatory policy argument, and* treats a *proposed GDPR* provision on automated and machine-readable privacy management *as a policy* case *for* moving from banner-based compliance toward *rights-supporting digital infrastructure.*

## Introduction

The internet is often narrated as an engineering triumph: a layered, interoperable network that made communication cheap, scalable, and generative. That story is accurate but incomplete. The same architecture that enabled permissionless innovation did not provide a native, institutionally complete capability through which people could express, remember, withdraw, delegate, or contest rights-relevant choices in a reliable way. Privacy, consent, objection, and other rights were therefore externalized to websites, platforms, browsers, and later consent-management vendors. The result is familiar to almost every internet user: repeated banners, opaque privacy policies, asymmetric buttons, fragmented records, and a persistent gap between formal rights and practical control (Acquisti et al. 2015; Nouwens et al. 2020; Solove 2013; Utz et al. 2019).

This paper advances a policy and regulation argument for information systems (IS): digital rights must be embedded in internet and software architecture, not merely declared in statutes or delegated to service-specific interfaces. A rights layer should enable standardized, machine-readable, bidirectional, and jurisdictionally plural communication among data subjects, controllers, processors, user agents, trusted intermediaries, and regulators without deciding substantive law. Architecture cannot generate political will, honest implementation, adjudication, or sanctions. It can redistribute initiation, representation, memory, limited evidence, coordination, and observability away from controller-controlled channels; law and institutions determine when communicated acts bind and what follows from obstruction.

The policy problem has become acute because major jurisdictions follow different logics. California and several other United States privacy frameworks are primarily associated with opt-out rights such as refusing the sale or sharing of personal information. The California Department of Justice recognizes Global Privacy Control (GPC) as a user-enabled "opt-out preference signal" that covered businesses must honor as a request to opt out of the sale or sharing of personal information, and the W3C GPC draft defines a signal for requesting that websites and services not sell or share personal information with third parties (California Department of Justice n.d.; W3C Privacy Working Group 2026). By contrast, the European Union (EU) combines multiple lawful bases under the General Data Protection Regulation (GDPR), including consent, contract, legal obligation, public tasks, and legitimate interests, with withdrawal and objection rights (GDPR arts. 6, 7, and 21); ePrivacy Directive art. 5(3), as amended, requires consent for storing or accessing information on terminal equipment unless an exception applies (Hoofnagle et al. 2019). The EU is therefore not a simple opt-in order, but it is certainly not reducible to a binary opt-out order.

This difference matters because digital architecture travels. Browser vendors, large platforms, advertising infrastructures, and standardization bodies can make one jurisdiction's technical grammar appear natural everywhere else. If a signal designed around California's opt-out logic becomes the default template for 'global' privacy communication, other legal orders may be forced to adapt their rights to a thinner technical form. This is not a problem with GPC itself. GPC is a noteworthy advance within the legal setting it was designed to serve. The danger is jurisdictional overreach through infrastructure: a binary signal that fits one law may become, through platform power and convenience, the practical boundary of rights in very different legal systems.

The European Commission's Digital Omnibus proposal (COM(2025) 837 final; 2025/0360(COD)) placed that architectural possibility on the legislative agenda through a proposed GDPR provision initially numbered Article 88b. Its significance here lies in the design problem it exposes, not in the provision's eventual numbering, wording, or legal fate. The proposal would require online interfaces to support automated and machine-readable indications of consent, refusal, and objection. The European Data Protection Board (EDPB) and European Data Protection Supervisor (EDPS) support addressing consent fatigue through machine-readable choices while recommending legal and technical clarifications and safeguards (EDPB and EDPS 2026). A rights-first and law-first implementation through a mechanism such as Advanced Digital Protection Control (ADPC) could help move online privacy from repetitive notice-and-click rituals toward a genuine rights communication infrastructure (Human 2026).

The primary research question is: *What explains the persistent gap between formally recognized [digital] rights and their practical exercisability when those rights are mediated through information systems under heterogeneous regulatory regimes?* The secondary research question is: *What normative requirements follow from this explanation for a human-compatible sociotechnical response that narrows the gap while preserving legal plurality and avoiding solutions that entrench controller power or merely shift it to client-side intermediaries?* Privacy and data protection provide the developed EU–California and Article 88b/ADPC case; the framework may extend to other rights with context-specific adaptation. Contributions are (1) *a* relational critical explanation of how legal heterogeneity, incentives, sociotechnical mediation, and uneven institutional capacity allocate power and reproduce the gap; (2) the governed rights-layer capability and its redistribution-or-reproduction mechanism; and (3) seven normative requirements (NR1–NR7). Operations and the lifecycle test coherence; empirical and design evaluation forms the natural next stage, building on the critical IS account and normative requirements developed here.

## Research Approach and Critical IS Contribution

This critical-conceptual IS policy inquiry examines how digital systems shape rights exercisability. It rests on the normative position that people should be able to exercise, understand, remember, delegate, withdraw, and contest rights in digital contexts through channels not controlled exclusively by actors with conflicting interests. Following Myers and Klein (2011), critical IS governs the inquiry; privacy theory addresses subject formation, information-infrastructure and internet-governance research address evolution and contestation, and bounded sociomateriality addresses enactment. This consequential problem framing and visible contribution follow IS guidance (Gupta 2018; Rai 2017).

The inquiry joins critique to a bounded transformative orientation. Its critical concepts are rights exercisability and procedural, epistemic, and infrastructural power: who can enact formal entitlements, control rights-relevant acts, define and retain categories and evidence, and set participation conditions. They challenge controller-controlled banners and isolated clicks as adequate enactments of rights, drawing on accounts of code, infrastructure, surveillance capitalism, and data colonialism (Couldry and Mejias 2019; DeNardis 2012; Lessig 1999; Zuboff 2015). Accordingly, emancipation means strengthening people's capacity to express, remember, withdraw, delegate, and contest rights; technology alone is insufficient. This requires reform across law, standards, organizational routines, user-side systems, oversight, and remedies.

Scholarly, legal, and standards materials were selected purposively for relevance to rights exercisability, jurisdictional variation, protocol and standards history, and IS theories of critique, infrastructure, and governance. Six linked steps were followed: (1) frame the phenomenon as a failure of rights exercise rather than banner usability; (2) state the normative position and challenged practice; (3) map peer-reviewed research against primary legal and standards documents; (4) compare California opt-out, the EU's mixed lawful-basis regime, and broader plurality; (5) reconstruct P3P, DNT, GPC, and ADPC sociotechnically and examine DPV and ODRL as complementary installed-base resources; and (6) combine the critical concepts with infrastructure and internet-governance theory to explain the gap (RQ1) and derive the capability, NR1–NR7, risks, and evaluation paths (RQ2). The traceable hierarchy is evidence and theory → diagnosis → explicit critical or normative inference (meta-requirement) → NR1–NR7 → illustrative operation → future implementation feature. Synthesis and problematization operate as methods within the critical inquiry (Alvesson and Sandberg 2011; Jaakkola 2020).

Rigor is pursued through visible values, conceptual consistency, evidentiary fit, power disclosure, inclusion, reflexivity, practical consequence, and explicit failure conditions. Interacting mechanisms, not universal effects, are identified. Peer-reviewed research supports empirical claims and primary legal or standards materials support institutional claims; clarity, scope, coherence, boundedness, and generativity assess conceptual adequacy (Gregor 2006; Gupta 2018; Whetten 1989). The mainly EU–California machine-readable case may underrepresent rights, practices, and communities outside those regimes or resistant to codification. The proposition and requirements remain fallible and contestable through empirical inquiry, institutional experience, and evaluation by affected communities.

## Critical-Theoretical Lens: Rights, Power, and Sociotechnical Infrastructure

The gap persists when heterogeneous entitlements lack interoperable channels; controllers and client providers have incentives to retain control; fragmented interfaces, standards, records, and workflows asymmetrically allocate initiation, representation, memory, evidence, and contestation; inaccessible channels externalize cognitive and social burdens; and weak integration, oversight, or consequence makes obstruction cheap and opaque. Privacy, the developed case, is relational: autonomy and subject formation depend on institutional and operational conditions, not isolated choices (Cohen 2000, 2019). These relations allocate procedural power over action, epistemic power over categories and records, and infrastructural power over participation (Cecez-Kecmanovic et al. 2014; DeNardis 2012). A rights layer has bounded emancipatory potential to redistribute initiation, representation, memory, limited evidence, coordination, and observability. Yet power can reappear in client defaults, controller workflows, standards, or investigation. Architecture makes exercise and obstruction communicable; law and institutions supply recognition, adjudication, and consequence. Bounded sociomateriality clarifies these relations (Rieskamp 2026), and this conditional redistribution-or-reproduction mechanism generates the governed capability below.

Internet governance likewise shows that code, markets, law, norms, and standards interact and that standards are sites of political struggle (DeNardis 2012; Lessig 1999). The TCP/IP reference model distinguishes application, transport, internet, and link functions, although protocols interact and functions may span layers (Braden 1989). Routing and naming are therefore treated as protocol-enabled functions and security as cross-cutting, not as separate layers. 'Rights layer' is an analytical and architectural concept for this governed capability, not a protocol-stack stratum. Although HTTP is stateless (Fielding et al. 2022), rights-related state can be maintained through linked messages, identifiers, acknowledgements, receipts, client records, and organizational workflows; their fragmentation is part of the problem. Existing standards make some policy and rights communication machine-readable, but no unified, broadly adopted, institutionally complete capability connects legally plural semantics, client-controller exchange, durable state, organizational handling, oversight, and remedy. More globally accepted arrangements may evolve; this analysis identifies what they must connect.

An information infrastructure is a shared, open, heterogeneous, evolving, recursively composed sociotechnical system of digital capabilities under distributed and emergent control (Hanseth and Lyytinen 2010). Within it, a *rights layer* is a focal governed sociotechnical capability for representing, communicating, retaining, updating, and contesting rights-relevant acts across services and regimes. The unit of analysis is this capability, produced by interoperable vocabularies and protocols; clients and trusted intermediaries; controller and processor systems; rights records and organizational routines; standards governance; oversight; and complaint channels. No actor owns it. Authorities establish or interpret rights and recognition duties; standards bodies steward semantics and profiles; clients, intermediaries, controllers, and processors implement components; independent bodies test conformance; civil society monitors; and regulators and courts investigate, adjudicate, enforce, and provide remedies. Any protocol, client, interface, record system, or process is only a component. The capability supplies interoperable conventions for choices, requests, reasons, acknowledgements, records, withdrawals, objections, and contestation; substantive law supplies the communicated rights and duties.

This framing clarifies the limits of *privacy by design*: GDPR art. 25 places duties principally on controllers but does not specify interoperable user-side rights channels. IS control and platform research shows that configured controls differ from enactment and that architecture and governance allocate decision rights (Tiwana et al. 2010; Wiener et al. 2016). The narrower contribution is whether a person or authorized agent can initiate and retain a legally situated act outside a duty-bearer's interface. Initiation, presentation, assistance, preference storage, and limited evidence may shift; processing control, legal effect, adjudication, and remedy do not, and clients or intermediaries may capture the shifted capacity. Privacy-enhancing technologies provide counter-power through prevention, minimization, blocking, and anonymity (Heurix et al. 2015), while intermediaries and assistants support decisions (Human, Alt, et al. 2022; Lehtiniemi and Kortesniemi 2017). A rights layer complements them through bidirectional communication of legally situated acts across services without substituting for controller duties, PETs, law, or enforcement.

‘Digital’ *protection* broadens the frame beyond ‘*data*’ *protection*, which remains central. Digital harms concern agency, manipulation, discrimination, collective autonomy, and democratic resilience: surveillance capitalism, hypernudging, and AI-enabled persuasion use data to shape attention, choice, and behavior (Susser et al. 2019; Yeung 2017; Zuboff 2015), while generative AI scales hyper-personalized and potentially epistemically invasive persuasion (LeBrun 2025). Do-*not-sell*/*share* categories—as exemplified by GPC—are therefore important but insufficient where harms arise from first-party processing, inference, and behavioral modulation without third-party transfer. *Digital protection* asks how environments can remain compatible with human rights, dignity, agency, and social sustainability. Privacy and data protection provide the developed case, not the boundary, of the rights-by-architecture claim. Generative AI, autonomous agents, robotics, other automated systems, IoT, and immersive technologies distribute rights-relevant interactions across physical and digital settings, showing why the capability must evolve beyond browsers, screens, and privacy (Hanseth and Lyytinen 2010; LeBrun 2025; Zheng et al. 2018).

## Why Notice-and-Choice Keeps Failing

The notice-and-choice model assumes that individuals can evaluate information, form preferences, express decisions, and later revise them. The empirical record repeatedly undermines that assumption. Privacy policies are often long, complex, unread, or poorly understood (Obar and Oeldorf-Hirsch 2020). Consent banners can use salience, defaults, asymmetry, hidden settings, and other dark patterns to steer acceptance (Gray et al. 2021; Nouwens et al. 2020); large-scale evidence from commercial interfaces shows that such steering techniques extend beyond banners (Mathur et al. 2019). Even without overt manipulation, repeated, context-dependent choices tax attention and weaken meaningful engagement (Acquisti et al. 2015; Utz et al. 2019).

The deeper issue is not that users do not care about privacy. The *privacy paradox* is often misread as indifference, but more careful analysis shows that constrained contexts, limited alternatives, social pressures, and cognitive burdens shape behavior (Solove 2021). In other words, observed clicks are poor evidence of unconstrained preference. A system that asks individuals to repeatedly manage complex legal and technical trade-offs without durable memory, human support, user-side assistance, or usable records is not empowering them; it is outsourcing systemic complexity to the least resourced actor in the chain.

The current model also fails temporally. Consent, refusal, withdrawal, and objection are continuing relations. A person may lack a portable, intelligible record or cross-service view of what was communicated, to whom, under which description, and when. GDPR art. 7(3) recognizes withdrawal and requires it to be as easy as giving consent, but rights remain difficult to manage when exercise is distributed across service-specific channels (Solove 2013). The paper therefore diagnoses an evidentiary asymmetry wherever a duty-bearer retains the operative state while the person lacks a corresponding record across services, devices, and time.

The model fails collectively as well. Privacy decisions can affect other people, as when contact lists, group photographs, household sensor data, or genetic information reveal facts about relatives, friends, coworkers, or communities (Human 2024). Children require support, not solitary legal clicking, and online data privacy can be understood as part of children's media rights (Brown and Pecora 2014). Privacy, in these contexts, is not only individual self-management but relational and collective protection (Nissenbaum 2004; Human 2024). A rights layer should therefore allow trusted assistance and delegated or guardian-mediated support without turning that support into a new surveillance channel.

Finally, screen-only, controller-specific mechanisms can be insufficient in repeated, ambient, inaccessible, cross-device, or cross-service settings. Smart-home IoT devices can make data practices and privacy risks difficult to observe or verify (Zheng et al. 2018). The same interface limitation can arise in mixed-reality environments, robots, smart vehicles, wearables, and agentic AI systems, where a banner may be unusable. Machine-readable communication is one proposed support for such settings, not a universal replacement for intelligible and accessible human-facing interaction. Digital regulation cannot depend exclusively on an interface pattern that already fails on ordinary websites.

## Jurisdictional Mismatch: Opt-Out Signals and Mixed Rights Orders

California's privacy framework gives GPC a legitimate role. The Department of Justice describes it as a stop-selling-or-sharing switch that covered businesses must honor as a valid opt-out request (California Department of Justice n.d.). This regulatory advance shows that technical signals gain force through law and can spare people repeated site-by-site expression of the same recognized choice.

The EU grammar is richer. GDPR arts. 6, 9, and 21 combine multiple lawful bases, special-category protections, and context-dependent objection rights; ePrivacy Directive art. 5(3) generally requires consent for terminal-equipment access. Under GDPR arts. 4(11) and 7(3), consent must be freely given, specific, informed, and unambiguous, and as easy to withdraw as to give (European Data Protection Board 2020). One binary global signal cannot represent this grammar without loss.

**Table 1. Jurisdictional Logic and Communication Requirements**

| Regulatory logic | Typical user act | Communication requirement | Risk if flattened |
|---|---|---|---|
| California-style opt-out | Do not sell/share personal information. | A recognized global opt-out signal can be effective when legally enforced. | If exported globally, it may reduce richer rights to a binary refusal. |
| EU mixed lawful-basis framework | Consent, refusal, withdrawal, objection, and other rights depending on legal basis and purpose. | Bidirectional, purpose-specific, revocable, and auditable communication is needed. | A binary signal cannot capture purpose-specific acts or lawful-basis plurality. |
| Children and vulnerable-user regimes | Guardian support, age-appropriate information, heightened protection, and limits on profiling. | Delegation, support, accessibility, and contextual rules must be expressible. | One-size-fits-all signals ignore vulnerability and social support. |
| Future digital-protection regulatory regimes | Controls over personalization, manipulation, AI explanation, and ambient sensing. | Extensible vocabularies and machine-readable rights acts are needed. | Legacy privacy signals may become obsolete or misleading. |

This difference shapes what a rights layer must communicate. A California opt-out can express no sale or sharing; an EU channel may need to carry a purpose-specific consent request, another lawful basis, refusal or objection, and evidence of later withdrawal. That requires vocabulary, context, bidirectionality, and records. A binary signal can be one message within the layer, not the layer itself.

Technology concentration amplifies the mismatch. A few dominant firms shape browsers, advertising systems, app ecosystems, and cloud infrastructure; standardization thus becomes a tussle in which architecture advances interests (Clark et al. 2005). Letting the easiest technical solution define rights can weaken legal plurality through protocol design rather than formal repeal.

This creates a risk of *jurisdictional digital imperialism*: infrastructures can extend asymmetrical power through extraction, dependency, and externally defined technical orders (Couldry and Mejias 2019; Kwet 2019). A standard built for a dominant market may become the de facto global form of rights exercise. A human-compatible layer must allow plural legal grammars rather than let one binary signal define the bounds of global digital rights.

Standards remain necessary, but should carry legal plurality: a rights layer must accommodate GPC-like opt-outs and ADPC-like consent, refusal, withdrawal, and objection without universalizing either regime. Table 1 outlines selected regulatory logics and their communication implications.

## Protocol Histories and Lessons for Regulation

Protocol history shows that technical means, legal mandates, and market incentives must align. P3P made website practices machine-readable so browsers could compare them with user preferences, but depended on accurate self-description and saw limited deployment and misused compact policies (Cranor et al. 2008; Leon et al. 2010). Technical machine-readability was feasible; *self-declared transparency without* incentives *and* enforcement was *fragile*.

DNT let browsers send a *do-not-track* preference and sites return status, but compliant behavior remained outside the specification. After insufficient deployment, the W3C process ended in a 2019 Working Group Note rather than a Recommendation (W3C Tracking Protection Working Group 2019).

GPC connected a browser signal to a California right: California recognizes it for sale/share opt-outs, demonstrating that browsers can communicate rights firms must honor (California Department of Justice n.d.). Its scope remains intentionally narrow; the W3C draft defines a sale/sharing signal, not a vocabulary for consent, withdrawal, objection, and contextual rights (W3C Privacy Working Group 2026).

ADPC takes a richer direction: a public specification for bidirectional communication of data-protection information, requests, preferences, and decisions between user-side software and services (Human et al. 2021). It supports general and specific choices, consent, refusal, withdrawal, and objection and may support compatible legal profiles. ADPC currently specifies communication paths using HTTP, JavaScript, and Bluetooth, illustrating how one component of the broader rights layer may operate through different technical channels. The broader rights layer also includes vocabularies, client and controller processing, records, workflows, governance, oversight, and remedies; it is neither identical to ADPC nor confined to the web.

Table 2 summarizes the sequence: P3P exposes unenforced self-description; DNT, simplicity without institutional force; GPC, recognized but narrow opt-out; and ADPC, richer bidirectional communication. These lessons orient Article 88b toward legally grounded automation that keeps supported and vulnerable people involved.

**Table 2. Protocol Lessons for a Rights Layer**

| Mechanism | Core idea | Key lesson for a Rights Layer |
|---|---|---|
| P3P | Machine-readable privacy-policy descriptions. | Semantic structure matters, but self-description without incentives and enforcement is weak. |
| DNT | A browser signal expressing a do-not-track preference. | Signals without legal and institutional force are easily ignored. |
| GPC | A legally recognized opt-out signal for sale or sharing in regimes such as California. | Legal recognition can make a signal effective, but binary opt-out logic is too narrow for frameworks such as the EU's. |
| ADPC | A bidirectional channel for consent, refusal, withdrawal, objection, requests, and records. | A rights layer for mixed regulatory regimes must be richer than one global flag and capable of user-side support. |

## Proposed Article 88b and the EU Opportunity

The proposed provision initially numbered GDPR art. 88b (COM(2025) 837 final) serves as the worked case for automated, machine-readable choices. Implementation would depend on standards, clients, controllers, oversight, and enforcement. The EDPB and EDPS support addressing consent fatigue through such choices while seeking legal and technical safeguards and clarifications (EDPB and EDPS 2026).

The illustrative Article 88b lifecycle tests how a standardized channel can carry rights-relevant acts over time: A purpose-specific controller proposition may reach a chosen client, or a person may initiate a request through it; ADPC messages are one possible exchange component (Human et al. 2021; Human 2026). Cognitive, contextual, accessibility, or authorized collective support can inform consent, refusal, objection, or another request. The controller acknowledges receipt and status, links the act to its workflow, and returns a minimal receipt supporting later update or withdrawal. An absent or contradictory response can enter a complaint or audit pathway; only a competent institution determines breach, remedy, or sanction. These roles may generalize to other digital rights, but their legal meaning and recognition duties do not automatically generalize.

The EU has distinctive reasons to act: Charter arts. 7–8 protect private life and personal-data protection, ECHR art. 8 protects private and family life, and the GDPR is one operational expression of related commitments. Those rights remain difficult to realize through manipulative, repetitive, controller-controlled interfaces. If consent enacts autonomy and dignity, its infrastructure must be compatible with people and legal requirements for validity. Article 88b could give future digital rules shared communication for requests, decisions, and records, but adoption must remain disaggregated. User-side assistance may be chosen; recognition is mandatory only where law requires it. Client conformance, controller processing, standards participation, investigation, adjudication, and remedy remain separate institutional questions; the capability coordinates them without one owner or universal duty. Article 88b must not become blanket browser-default consent or a channel limited to dominant browsers. Standardization should include supervisory authorities, civil society, accessibility and child-protection experts, smaller firms, technical experts, and diverse jurisdictions. It must be deliberative enough for legitimacy but timely enough that incumbents cannot indefinitely preserve banner dependence.

## Normative Requirements for a Human-Compatible Rights Layer

The analysis identifies seven contestable normative requirements. ***First (NR1: legal plurality), the capability must be legally plural.*** It should carry California opt-outs, EU consent and objection, and future rights without privileging Californian, EU, or advertising-industry ontologies or presuming a privileged adult as the default rights-holder. This requires extensible vocabularies, accountable governance of semantics, and NR5's human-compatible support. Technical neutrality is not enough; legal pluralism must be built into the institutional and technical arrangement (Hoofnagle et al. 2019; Human et al. 2021).

***Second (NR2: bidirectionality), the layer must be bidirectional.*** Controllers need to communicate requests, purposes, legal bases, consequences, and changes. Data subjects or their agents need to communicate decisions, refusals, withdrawals, objections, questions, and preferences. A one-way signal may be useful for simple opt-outs, but it cannot support the communicative richness of EU data protection or the contextual complexity of pervasive digital environments (Human, Pandit, et al. 2022).

***Third (NR3: person-held records), the layer must support durable user-side records.*** People should be able to inspect what was requested, what was decided, when it was decided, and how to revise it. Records should not become new tracking identifiers, and privacy-preserving design is essential. Yet without records, withdrawal and accountability remain dependent on controller-controlled logs. User-side receipts can rebalance evidence without centralizing everything in state or platform databases.

***Fourth (NR4: accountable client mediation), the layer must shift interface power.*** If controllers define every button, color, default, and pathway, dark patterns will persist. User-side software can present requests through standardized, accessible, and independently audited interfaces. This does not eliminate design problems; browsers and operating systems can also manipulate. It shifts control over initiation and presentation and may improve observability, while controller incentives remain an institutional problem. User-side interfaces can also support multiple languages, accessibility settings, plain-language explanations, and visual or audio modalities.

***Fifth (NR5: human-compatible support), the layer must enable cognitive, contextual, and collective support.*** Cognitive support includes simplification, automation, memory, summaries, and warnings. Contextual support allows different choices for different purposes, devices, times, or risk levels. Collective support allows children, elderly users, disabled users, or anyone who chooses assistance to involve guardians, experts, civil-society organizations, or trusted communities. Such support need not be paternalistic when it is transparent, revocable, and chosen or legally authorized. It is recognition that humans are social and limited, not isolated preference machines (Human and Cech 2021; Human 2024).

***Sixth (NR6: auditable handling), use of the capability must be connected to enforceable duties and be auditable.*** Regulators need to know whether a controller recognized a signal, whether a request matched a standard vocabulary, whether consent was validly obtained, and whether refusal or withdrawal was honored. Machine-readable communication can make these questions easier to audit, but only if standards require verifiable traces and if sanctions make noncompliance costly. DNT's insufficient deployment shows the risk when recognition and consequence remain external to a specification. Rights-layer implementations—including those envisaged by proposed Article 88b—should avoid that mistake.

***Seventh (NR7: adaptability), the layer must be future-facing.*** First-party data use, generative AI, hyper-personalization, mixed reality, robots, and IoT show that digital protection cannot be reduced to third-party tracking. A person may not be concerned only that data is sold or shared; they may be concerned that data is used to manipulate, infer vulnerability, profile a child, personalize prices, produce political persuasion, or adapt an AI companion. A rights layer should therefore be designed as a foundation for digital protection, not only for cookie consent (Human 2024; LeBrun 2025; Susser et al. 2019; Yeung 2017).

Figure 1 provides an illustrative, non-exhaustive mapping for NR1–NR7, linking each diagnosis and meta-requirement to selected actors, capture risks, and prospective indicators.

**Figure 1. Illustrative derivation mapping for the seven normative requirements.**

| SELECTED DIAGNOSIS | META-REQUIREMENT | NR | SELECTED ACTOR ROLES | ILLUSTRATIVE RISK | PROSPECTIVE INDICATOR |
|---|---|---|---|---|---|
| Plural legal meanings / fragmented channels → | Preserve legal plurality → | NR1 | Authorities · standards bodies · implementers | Dominant ontology | Profile coverage · mapping accuracy |
| One-way communication → | Reciprocal communication → | NR2 | Clients · controllers | Selective or misleading acknowledgment | Request–response completeness |
| Asymmetric record control → | Rebalance evidence → | NR3 | Clients · controllers | Tracking or mismatched records | Receipt portability · reconciliation accuracy |
| Controller-controlled interfaces → | Accountable mediation → | NR4 | Client providers · auditors | Client capture | Independent conformance · switching feasibility |
| Cognitive and social burden → | Bounded support → | NR5 | Clients · intermediaries · communities | Paternalism or exclusion | Accessibility · revocability · experienced burden |
| Opaque or low-cost obstruction → | Observable handling and consequence → | NR6 | Implementers · testers · regulators · courts | Ignored acts · weak consequence | Acknowledgment integrity · auditability · remedy access |
| Evolving rights and digital environments → | Maintain adaptability → | NR7 | Standards bodies · authorities · implementers · communities | Lock-in · outdated profiles | Gateway interoperability · revision responsiveness · new-domain coverage |

Arrows show an illustrative derivation path; the remaining columns identify selected roles, risks, and future indicators rather than exhaustive causal steps.

## Rights-Layer Operations: What Must Be Communicated

Six illustrative operations (Table 3) show how NR1–NR7 apply across actors and time: information or request, decision, record, update or withdrawal, delegated or supported action, and audit trace. They are analytical categories, not findings, wire commands, or a universal sequence; ordering depends on the applicable right and regime. The layer should carry controller-provided information or propositions, person-initiated requests, context, decisions (including objections), legally relevant reasons, updates or withdrawals, the scope of assistance or delegated authority, and acknowledgments, while supporting minimal records and audit evidence. Together, the operations establish context, express and link rights-relevant acts, revise prior acts, enable assistance, and support review.

This operational view matters in the EU because GDPR arts. 7 and 12–22 distinguish consent and withdrawal, controller-provided information, person-initiated requests, objections subject to different conditions, and rights concerning qualifying automated decisions. Timing and effects vary: information may precede processing, requests trigger responses, some objections require grounds but direct-marketing objections do not, and withdrawal does not affect prior lawful consent-based processing. A thin signal cannot carry these distinctions. The rights layer should therefore be a governed communication capability for rights-relevant acts; exchanges may be protocol-mediated, but the capability is neither a symbolic preference nor a protocol alone.

The operational view also changes compliance evaluation. Cookie-banner audits often reconstruct interfaces from screenshots: which buttons appeared, what colors were used, how many clicks were required, and whether cookies were set before consent (Santos et al. 2020; Utz et al. 2019). These methods remain useful but laborious and reactive. Audit traces could enable direct inspection of how rights-relevant acts were handled. Regulators could ask whether controller-provided information accurately stated the relevant purpose and legal basis; whether a person-initiated request or decision was received and handled as required; whether later processing remained consistent with that act and applicable law; and whether an update or withdrawal was honored.

For organizations, the operations support a governed workflow rather than a collection of visual widgets, aligning privacy engineering with IS concerns about process governance, interorganizational systems, and accountability. For users, records provide continuity: what matters may not be the initial click but the later ability to know and change what happened. A user-side record should not become a surveillance dossier; it should be minimal, local where possible, secure, and user-controlled. Without it, a person may lack durable evidence of withdrawal, leaving accountability largely to the controller's version of events. The aim is reduced evidentiary asymmetry—not data accumulation.

Finally, these operations support contestation. Digital services often make decisions through models, recommender systems, and personalization engines whose effects are difficult to perceive. By linking decisions—including objections—and explanations to records and audit traces, the rights layer can support challenges not only to data collection but also to uses incompatible with agency, fairness, or legal limits. This makes the move from '*data*' *protection* to '*digital*' *protection* operational rather than rhetorical. Communicating rights-relevant acts is a first step toward contestable digital environments.

**Table 3. Rights-Layer Operations and Implementation Implications**

| Operation | Why it matters | Implementation implication |
|---|---|---|
| Information or request | Controller-provided information or propositions supply the purpose, legal basis, actors, and context; person-initiated requests convey the action sought and its scope. | Use standardized descriptors for each direction and jurisdiction-specific vocabularies. |
| Decision | Communicates consent, refusal, objection, or another legally relevant act. | Represent legally distinct acts rather than a single binary flag. |
| Record | Supports later withdrawal, auditing, and accountability. | Store minimal user-side receipts securely without creating stable cross-site identifiers. |
| Update or withdrawal | Recognizes that rights exercise continues beyond a single click. | Make withdrawal as easy as consent; make later changes machine-readable where applicable. |
| Delegated or supported action | Enables trusted people or assistive tools to help while preserving autonomy. | Makes rights exercise human-compatible; requires transparency, revocability, and clear authority boundaries. |
| Audit trace | Allows regulators, researchers, or courts to assess whether rights-relevant acts were handled as required. | Create evidence hooks for compliance testing while minimizing disclosure. |

Table 3 summarizes six illustrative rights-layer operations, why they matter, and their implementation implications; the following crosswalk links each operation to selected NRs, actors, evidence, dependencies, and failure risks: Information or request (NR1–2) may originate with a controller or person and pass through a client. Controller-provided information or a proposition supplies the purpose, legal basis, actors, and context; a person-initiated request conveys the action sought, scope, actors, and context. Shared vocabularies are necessary; missing, inaccurate, or misleading context causes failure. A decision (NR1–2, 4–6) conveys consent, refusal, objection, or another legally relevant act; a linked record should capture receipt and status. Law and competent institutions determine legal effect; conformance and organizational handling determine implementation, while coercion and misleading status remain risks. A record (NR3, 6) gives relevant parties minimal, reconcilable receipts but risks insecure storage, mismatch, or tracking. An update or withdrawal (NR2–3, 6) links a revised or revoked act to its receipt but fails if organizational processes ignore it. Delegated or supported action (NR4–5) requires visible, bounded, revocable assistance or authority; capture remains a risk. An audit trace (NR3, 6) provides verifiable evidence, but accountability also requires accessible review or complaint routes and consequences. NR7 is cross-cutting: operations, vocabularies, and profiles must adapt to changing legal, technological, and social conditions while preserving earlier acts' interpretation. Together, the operations form the paper's illustrative rights-layer scenario. The proposed Article 88b use case, with ADPC as one possible exchange component, illustrates rather than prescribes a sequence.

## Human-Compatible Automation and Delegation

An objection to automated rights communication is that it may hide decisions and undermine consent. The concern matters, but current practice already involves inattentive, repetitive, and often steered clicking (Nouwens et al. 2020; Utz et al. 2019); manual action is not necessarily meaningful action. The question is how support can reduce burden while preserving agency, reviewability, and contestation. Morel and Fischer-Hübner (2023) classify *automation* levels and associated challenges; Andreotta (2025) examines informed-consent implications; and Morel et al. (2025) map privacy-assistant approaches and evidence gaps. Together with work on consenting assistants and collective support, these sources warrant configurable and reviewable assistance rather than displacement of human judgment (Lehtiniemi and Kortesniemi 2017; Human, Alt, et al. 2022; Human 2024; Human 2026).

Two implications of NR3–NR5 *are configurability and periodic review*. Users should be able to choose different levels of automation, from reminders and contextual prompts to refusal rules, expert-recommended lists, child-protective settings, or accessibility-oriented explanations, and retain a manual path where law permits. The source, scope, and operation of assistance must be visible, intelligible, accessible, reviewable, configurable, and revocable; users should be able to refuse or contest it, and any override or route of appeal should be visible. Periodic summaries, risk alerts, and dashboards can return decisions to attention without recreating banner fatigue. The criterion is the quality of continuing agency, not simply fewer clicks. These are normative requirements; future design work can operationalize them through explicit design-principle anatomy and development steps before evaluating specific artifacts (Gregor et al. 2020; Möller et al. 2020).

Two further implications of NR3, NR5, and NR6 *are accountable delegation and privacy-preserving processing*. Parents, caregivers, professionals, and civil-society organizations may support decisions, but authority must be scoped, logged, and revocable, and supporters' access minimized. The critical questions are whose preferences are encoded, who can override them, and whether cognitive or administrative burdens are shifted onto users, caregivers, or marginalized groups. Local rules, downloadable vocabularies, and minimum-necessary signals can reduce data flows; otherwise a protection layer could become a surveillance layer. These requirements follow from collective privacy effects and the limits of individual self-management (Lehtiniemi and Kortesniemi 2017; Human 2024).

## Legal Pluralism Beyond the EU and California

The contrast between California and the EU is analytically useful, but a global rights layer must anticipate other jurisdictions and sectoral rules. Many legal systems combine privacy, consumer protection, children's rights, online safety, AI governance, and competition concerns. Some emphasize opt-out rights, others prior consent, and still others duties of fairness, transparency, or authorization. A layer built around one legal grammar would become an infrastructural bottleneck for every jurisdiction whose rights do not fit its vocabulary. Juxtaposing comparative analysis of the GDPR with GPC's deliberately bounded opt-out scope yields the inference that technical standards cannot safely assume a single legal basis or rights act across jurisdictions (Hoofnagle et al. 2019; W3C Privacy Working Group 2026).

Plurality requires modular, profile-based evolution, not hard-coded national doctrine: a core for actors, acts, status, time, and revocation; jurisdictional vocabularies for bases, purposes, exceptions, and safeguards; and gateways to existing systems. Infrastructure constraints are installed-base cultivation, gateways, modularity, bootstrapping, adaptability, and backward compatibility (Hanseth and Lyytinen 2010); rights-specific additions are legal plurality, bounded assistance, accountable client mediation, person-held evidence, observable handling, and contestable consequence.

The installed base is not empty. ODRL models policies, permissions, prohibitions, duties, parties, actions, constraints, and extensible profiles (W3C Permissions & Obligations Expression Working Group 2018a, 2018b); DPV 2.3 represents processing and rights concepts, records, and statuses, although it is a Final Community Group Report, not a W3C Standard (W3C Data Privacy Vocabularies and Controls Community Group 2026). ADPC supplies bidirectional messages, ISO/IEC TS 27560:2023 consent records and receipts, and IEEE 7012-2025 machine-readable contractual privacy terms and party-held records (Human et al. 2021; IEEE Standards Association 2026; International Organization for Standardization and International Electrotechnical Commission 2023). None alone integrates plural public-law semantics, client–controller exchange, organizational handling, recognition, oversight, and remedy. The proposal connects them to P3P, DNT, GPC, browsers and other clients, controller systems, and complaint processes through profiles, gateways, testing, and revision; broader standards may emerge through cumulative use and contestation.

## Reviewability and Contestability

Reviewability elaborates NR3 and NR6. Users should be able to inspect summarized histories of important requests, see which decisions were automated, understand which rule or trusted list was applied, and revise decisions without reconstructing an entire browsing history. Reviewability is the safeguard that distinguishes support from hidden paternalism and makes automated rights exercise more defensible under legal regimes that require meaningful control (Human 2024). The analysis therefore treats rights that cannot be remembered, inspected, or revised as weak in practice; that inference follows from documented self-management burdens and communication-assistant design (Human et al. 2021; Human, Alt, et al. 2022; Solove 2013).

*Contestability* is the institutional condition that gives NR6 practical force. A user, regulator, auditor, or civil-society organization should be able to challenge whether a request was valid, whether a vocabulary term was used honestly, whether a refusal was ignored, or whether a later data use exceeded the communicated purpose. This is especially important for children, people with disabilities, and users who delegate decisions: logs of delegated action, revocation of delegated authority, plain-language explanations, and independent oversight should enhance human agency rather than replace one inaccessible interface with another opaque institution. It also connects to legal and technical auditability in consent-banner research, where compliance increasingly depends on reconstructing what was displayed, communicated, and acted upon (Santos et al. 2020; Utz et al. 2019).

## Digital Protection, Children, and Pervasive Environments

NR7 carries the requirements into broader digital environments. The move from ‘*data*’ *protection* to ‘*digital*’ *protection* is necessary because digital harms increasingly arise through uses, inferences, and environments, not merely through transfer. A system can respect a do-not-sell signal and still use first-party data to personalize content in ways that shape attention, emotions, and behavior. A system can avoid sharing data and still create manipulative recommendations, discriminatory inferences, or exploitative adaptive interfaces. *Digital protection* therefore treats data protection as necessary but incomplete. The object of concern is the protection of human agency, dignity, privacy, equality, and democratic self-determination in digital environments (Human 2024; Susser et al. 2019; Yeung 2017).

Children make the point vivid. A child cannot be expected to read legal notices, infer long-term consequences, and manage cross-service records. Yet children increasingly interact with games, learning platforms, social media, smart toys, voice assistants, and immersive environments. The rights layer proposed here would not simply ask children to click better. It would enable age-appropriate explanation, guardian-mediated decision support, protective defaults, and external oversight. Such support aligns with the view that online data privacy is part of children's media rights (Brown and Pecora 2014).

Vulnerability is not limited to children. Elderly users, disabled users, migrants, people with limited literacy, people under stress, and people in coercive relationships may all need different forms of support. A human-compatible layer should allow multilingual interfaces, screen-reader compatibility, simplified explanations, trusted intermediaries, and contextual warnings. This is a regulatory issue, not only a design preference. If rights are exercised only through interfaces that many people cannot use effectively, rights protection becomes socially unequal (Human 2024).

Pervasive environments intensify the problem. Smart homes and IoT devices create privacy concerns that users often cannot observe or control at the moment of data collection (Zheng et al. 2018). Mixed reality can track gaze, gestures, surroundings, bodies, and social interactions. Robots and automated systems may sense and act in shared spaces where bystanders did not initiate the interaction. In these settings, rights communication cannot depend on a single user's website visit. It must operate across devices, spaces, and affected persons.

Generative AI and hyper-personalization further shift the stakes. Consent to data processing may become a gateway to adaptive persuasion, synthetic content, personalized pricing, or emotional manipulation. Recent work on AI-generated hyper-personalized digital advertisements highlights risks that extend beyond ordinary targeted advertising (LeBrun 2025). A rights layer should therefore evolve toward communicating not only whether data is shared, but also whether and how it may be used for personalization, profiling, inference, or behavior-shaping. Digital protection is the umbrella under which such future rights can be made communicable.

## Standardization, Power, and the Risk of Capture

Standardization is often presented as a neutral process of technical consensus. In practice, it is a governance arena: standards allocate adjustment costs, defaults, and practical capabilities. Well-resourced technology firms, advertising networks, and client providers can participate continuously, while civil society, affected communities, and small firms may have less capacity. These resource asymmetries can allow dominant actors to shape standards around existing business models (Clark et al. 2005; DeNardis 2012; DeNardis and Hackl 2015). Standards should therefore be treated as contestable governance arrangements rather than neutral artifacts (Lessig 1999; DeNardis 2012).

These asymmetries create a specific capture risk for the rights-layer capability illustrated by proposed Article 88b. Even where law requires recognition, adverse power can reappear in implementation: a client may encode manipulative defaults; a controller may misclassify an act, return a misleading acknowledgment, or fail to connect it to operational workflows; a standards process may narrow the vocabulary; and an under-resourced authority may not investigate. Architecture is nevertheless consequential because a standardized act and status can make initiation, obstruction, and contradiction more observable. It redistributes some procedural capacity; it does not make compliance self-enforcing.

Adoption and honoring require different, jurisdiction-specific levers: binding recognition duties where law creates them; open standards, reference implementations, profiles, and gateways that lower implementation cost; public conformance tests and independent certification; procurement conditions for client and controller systems; accessible complaint and audit pathways; and liability or sanctions authorized by the relevant regime. Civil-society monitoring can expose patterns. Reputational or market pressure may be supplementary where observable practice and market conditions make it operative. Supervisory authorities, technical experts, accessibility and human-rights groups, small firms, and resourced public-interest actors should share stewardship; no lever applies identically to every right or jurisdiction.

There is also a governance risk on the user side. Browsers, operating systems, apps, extensions, assistants, accessibility tools, robots, and automated agents can relocate initiation, presentation, memory, and support, but can also impose manipulative defaults, proprietary vocabularies, sensitive state, concentration, misunderstanding, switching costs, and interoperability failures. People should be able to choose compatible clients and extensions where law permits. Multiple implementations, open profiles, portable records, legal and conformance duties on user-side actors, independent testing, and regulatory oversight can limit capture. Moving the interface must not replace controller dominance with client dominance or obscure effects on autonomy, accessibility, children, and collective interests.

Standardization delays have real consequences. DNT's conclusion as a 2019 W3C Working Group Note after insufficient deployment illustrates that technical specification without ecosystem adoption and institutional force may stall (W3C Tracking Protection Working Group 2019). A rights-adequate first version should therefore be pursued within explicit timelines and maintained as a living standard that evolves with legal, technological, and societal change.

## Evaluation Path for Future Systems

Future evaluation can test legal conformance (message completeness, recognition, and withdrawal), human consequences (understanding, accessibility, burden, and agency), and ecosystem effects (interoperability, adoption, and incumbent power). These are prospective criteria for specific implementations rather than reported outcomes (Gregor and Hevner 2013; Human, Pandit, et al. 2022). Field comparisons across websites, apps, IoT, and immersive systems; longitudinal receipt studies; behavioral audits after legally recognized signals; and standardization ethnographies could test adoption, inclusion, governance, and enforcement.

Success would mean better conditions of rights exercise, not elimination of digital harm, assessed through burden, dark-pattern exposure, withdrawal usability, accountability, accessibility, child and guardian support, compliance duplication, and auditability. The requirements are reasoned propositions for deliberation, profiling, and testing with standards communities and affected groups; further participatory and empirical work should refine them.

## Discussion and Implications for Information Systems

The primary research question is answered relationally: formal recognition alone does not reorganize rights-exercise systems. Legal heterogeneity, conflicting incentives, fragmented interfaces and records, human burdens, and uneven recognition, integration, oversight, and consequence preserve controller and client control, reproducing the procedural, epistemic, and infrastructural power explained above. Rights exercise is thus an IS *phenomenon* enacted through law, markets, interfaces, standards, routines, records, technologies, and institutions. Sociomateriality supports this relational account; infrastructure and internet-governance research explain distributed evolution and contestation (Cecez-Kecmanovic et al. 2014; DeNardis 2012; Hanseth and Lyytinen 2010; Star and Ruhleder 1996).

The hierarchy is (1) a critical account of practical digital-rights exercisability and power; (2) the governed capability and its redistribution-or-reproduction mechanism; and (3) NR1–NR7. The latter two answer RQ2 by specifying conditions for redistributing initiation, representation, memory, limited evidence, coordination, and observability; law and institutions retain authority over entitlements, legality, and consequence. Operations illustrate the requirements and the Article 88b lifecycle tests coherence; empirical and design evaluation forms the natural next stage. The requirements test whether a response narrows the gap without collapsing legal plurality or reproducing controller and client power. Authorities define scope; standards bodies steward profiles; clients and controllers implement; independent bodies test; regulators and courts enforce and remedy; civil society and affected communities govern; researchers evaluate consequences.

A subordinate implication is that future artifacts require more than click counts or banner acceptance rates: evaluation should combine legal validity, usability, accessibility, security, interoperability, organizational adoption, and market-power effects. IS research can develop personal protection assistants, guardian-mediated systems, privacy-preserving receipts, rights dashboards, and compliance tests while keeping values and stakeholder heterogeneity visible during construction (Gregor and Hevner 2013; Human, Alt, et al. 2022; Human, Pandit, et al. 2022).

Regulation-in-IS research distinguishes regulation of IS from regulation through IS while emphasizing fragmentation, cost, technological change, and more complex relationships (Väyrynen et al. 2025; Wurzer and vom Brocke 2025). *Rights by architecture* connects the two by specifying capabilities through which rights can be exercised. The proposal's initially numbered GDPR art. 88b (COM(2025) 837 final) would require machine-readable routes for consent, refusal, and objection. GPC fits sale/share opt-outs but should not define EU consent, withdrawal, objection, or lawful-basis plurality. ADPC or a similar mechanism is a stronger starting point because it communicates requests, decisions, revocations, objections, metadata, and records and can adapt across regimes (EDPB and EDPS 2026; Human et al. 2021; W3C Privacy Working Group 2026; Human 2026).

Machine-readable rights communication can support computational policy auditing and digital sustainability by giving regulators, researchers, civil society, and firms a shared object of inspection. Audits could test signal recognition, lawful vocabulary, refusal and withdrawal handling, and fingerprinting or delegation risks. A digital economy is not sustainable when its infrastructures normalize exhaustion, manipulation, opacity, or rights friction; *digital protection is therefore part of* IS *social and democratic sustainability* (DeNardis 2012; Human 2024; Santos et al. 2020; Utz et al. 2019; Zuboff 2015).

## Boundary Conditions and Risks

A rights layer cannot by itself remedy *surveillance capitalism*, platform concentration, manipulative business models, weak enforcement, extensive first-party use, sensitive inference, or pressured disclosure. Substantive regulation must address those practices. The layer may provide a necessary cross-service communication capability, but cannot itself supply recognition, legality, or remedy.

Automation can reproduce inattentive consent when broad defaults are set once and forgotten. Accountable automation instead requires periodic review, intelligible summaries, alerts for unusual requests, easy withdrawal, and trusted support. It should reduce cognitive burden while preserving agency (Human 2024; Lehtiniemi and Kortesniemi 2017).

User-side records may become sensitive repositories, signals may enable fingerprinting, and intermediaries may abuse delegated authority. Standards must minimize disclosure, avoid stable cross-site identifiers, favor local processing, and secure stored records so that protection does not create new surveillance (Human et al. 2021; W3C Privacy Working Group 2026).

Legal pluralism creates unavoidable n-to-m complexity. Shared act types, jurisdictional profiles, gateways, status conventions, and records may reduce repeated bilateral integration and expose differences; they cannot remove interpretation, exceptions, organizational change, transition, or contestation. Cognitive, contextual, and collective user-side support can reduce the complexity experienced by a person through tailored summaries, rules, warnings, assistive presentation, or chosen expert and community support without pretending that the underlying legal heterogeneity has disappeared. The capability communicates rights-relevant acts; it cannot override lawful authority, retention duties, or competent determinations of legality, necessity, proportionality, safeguards, oversight, and remedies.

Legitimacy is also at risk if large firms or technocratic institutions dominate standardization. Transparency, public consultation, civil-society participation, and independent evaluation are therefore necessary for governing this *constitutional-scale intervention* in digital life (DeNardis 2012; EDPB and EDPS 2026).

## Conclusion

Digital societies have placed rights into law without reorganizing rights-exercise systems. RQ1 is answered relationally: legal heterogeneity, adverse incentives, fragmented standards and workflows, unequal human support, and uneven oversight and remedy preserve controller and client control, concentrating procedural, epistemic, and infrastructural power. Architecture matters but is neither independently causal nor sufficient.

NR1–NR7 answer RQ2 through a human-compatible rights layer for standardized, machine-readable, bidirectional, and jurisdictionally plural communication of rights-related information, requests, decisions, and records. It should support California opt-out rights without universalizing them, EU consent and objection without reducing them to banners, and future rights without predetermining their substance. It should support vulnerable people, especially children, extend protection beyond data protection, and make rights communicable in IoT, mixed reality, robots, generative AI, and other automated systems where conventional interfaces fail.

The Commission proposal's provision initially numbered GDPR art. 88b illustrates one transition path. Implementation should neither universalize opt-out, create consent by default, empower only dominant browsers, nor let industry convenience define rights. A rights-first and law-first path should build on ADPC or a similar open, extensible mechanism linked to legal recognition, oversight, and enforcement.

For IS, policy and regulation are enacted through information systems and shaped by power in design and governance. Digital governance depends on making rights infrastructural without capture, opacity, or paternalism. Rights by architecture is a bounded response, not a complete solution; without it, protection remains trapped in the banner.